# Microwave shielding of transparent and conducting single-walled carbon nanotube films

Hua Xu and Steven M. Anlage
*Center for Superconductivity Research, Department of Physics, University of Maryland, College Park, Maryland 20742-4111*

Liangbing Hu and George Gruner[a]
*Department of Physics, University of California, Los Angeles, California 90095*

The authors measured the transport properties of single-walled carbon nanotube (SWCNT) films in the microwave frequency range from 10 MHz to 30 GHz by using the Corbino reflection technique from temperatures of 20–400 K. Based on the real and imaginary parts of the microwave conductivity, they calculated the shielding effectiveness for various film thicknesses. Shielding effectiveness of 43 dB at 10 MHz and 28 dB at 10 GHz are found for films with 90% optical transmittance, which suggests that SWCNT films are promising as a type of transparent microwave shielding material. By combining their data with those from the literature, the conductivity of SWCNT films was established in a broad frequency range from dc to visible. © *2007 American Institute of Physics*. [DOI: 10.1063/1.2734897]

Single-walled carbon nanotubes (SWCNTs) are emerging as building blocks of electronics for a variety of applications. In particular, films of nanotubes have found potential applications for electronics and optoelectronics.[1] However, an understanding of the transport mechanism that governs the response to dc and ac electric fields is required for application. It has been established that the overall resistance of the films is determined by the junction resistance between different tubes.[2,3] To fully explore the potential applications of NT films as a type of electrical and photonic material, it is important to map out the conductivity in a wide frequency range. Ac conductance measurements of a percolating NT network of up to 1 MHz have shown a universal power law in frequency, which is commonly observed in systems with randomly distributed barriers.[4] In the terahertz range, the effective Maxwell-Garnet model has been introduced where both the metallic and semiconducting NTs were embedded in a dielectric host.[5] Optical conductance in the far infrared and visible range has been obtained by measuring the reflectance of NT films and a Kramers-Kronig transformation.[6–8] Study of conductance in the microwave frequency range is important for high speed NT thin film field effect transistors. Microwave conductivity of individual SWCNT and operation as transistor at 2.6 GHz have been measured by Burke and co-workers.[9] They also gave a rf circuit model for carbon nanotubes.[10] However, there is a paucity of conductivity measurements on SWCNT films in this frequency range. So far, a few groups have measured with a cavity setup, which can only lead to conductivity values at a few discrete frequencies.[11,12] Here we use the Corbino reflectance setup[13] which allows the measurement of conductivity in a continuous frequency range from the rf to the microwave.

Electromagnetic radiation at radio and microwave frequencies, such as those emanated from cell phones, tend to interfere with electronic devices. The electromagnetic interference (EMI) leakage from radio frequency to microwave is still a serious problem for our society. The primary mechanism of EMI shielding is usually reflection. Thin metal foil and metal grids are commonly used for this purpose. Recently light-weight, flexible, and highly effective shielding can be achieved by means of a conducting polymer coating,[14] although degradation is an intrinsic problem. A matrix containing conductive fillers is an attractive alternative for shielding.[14] Composites incorporating NTs have extremely low percolation threshold (volume fraction around $10^{-4}$) due to the high ratio of length to diameter.[15] Highly conducting SWCNTs and multiwalled carbon nanotubes[16] have been incorporated into composites for EMI shielding purposes and a 49 dB shielding effectiveness at 10 MHz has been achieved for 15% NT loading of a polymer composite.[17]

In many situations visibly transparent EMI shielding is required and an indium tin oxide (ITO) is the current material of choice. NT films have shown potential as the replacement for ITO for transparent electrodes in devices.[1] Up to now, there is no systematic investigation on EMI shielding effects of NT films, especially for the microwave frequency range where the EMI leakage is mostly concentrated. In this letter, by measuring the microwave conductivity of NT films from 10 MHz to 30 GHz at different temperatures, we found that SWCNT films are promising for the application of transparent EMI shielding.

SWCNT films on polyethylene terephthalate (PET) substrates are fabricated by the transfer printing method.[18] A typical test film has 30 nm thickness and 80% optical transmittance, as shown in Fig. 1(a). A gold contact was deposited to form a Corbino disk on the sample surface, as shown in Fig. 1(b). Figure 1(c) shows the coaxial cable/film interface. With this setup, we are able to measure the complex reflection coefficient $\hat{S}_{11}$ of the film continuous over the frequency range from 10 MHz to 30 GHz using an Agilent E8364B PNA.[13] The surface impedance of the sample can be extracted from $\hat{S}_{11}$,

---

[a]Electronic mail: ggruner@ucla.edu

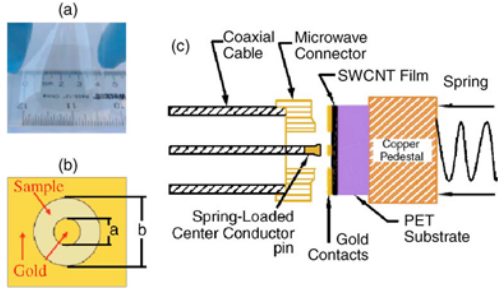

FIG. 1. (Color online) (a) Photograph of SWCNT films on PET substrate; (b) top view of the gold Corbino contact to the sample ($a=0.5$ mm and $b=1.9$ mm); and (c) interface of microwave connector to the sample (not to scale).

$$Z_s(\omega) = Z_0 \frac{2\pi}{\ln(b/a)} \frac{1+\hat{S}_{11}(\omega)}{1-\hat{S}_{11}(\omega)}, \quad (1)$$

where $Z_0 = 50\ \Omega$ is the characteristic impedance of the coaxial transmission line and $a$ and $b$ are the inner and outer diameters of the sample contact, respectively. The obtained surface impedance $Z_s(\omega)$ includes a contribution from the NT film $Z_s^{NT}$ and a contribution from the substrate and the sample holder $Z_s^{sub}$. To extract the surface impedance of the NT film we also measured $\hat{S}_{11}$ of the pure dielectric PET substrate on the same sample holder. In our measurement, the film thickness $t_0$ and the complex propagation constant $k_{NT}$ in the NT film satisfy $t_0|k_{NT}| \ll 1$ and measurements are made below the frequency of the lowest propagating waveguide mode in the sample/substrate so $[1/Z_s(\omega)] \cong [1/Z_s^{NT}] + [1/Z_s^{sub}(\omega)]$. The surface impedance $Z_s^{NT}$ of the NT film is obtained by subtraction and the conductivity is obtained as $\sigma(\omega) = 1/Z_s^{NT} t_0$.

Real and imaginary parts of the conductivity versus frequency are plotted in Figs. 2(a) and 2(b). For all the different temperatures, the conductivities keep their dc value up to about 100 MHz and start to increase at higher frequency. Both the real and imaginary parts of the conductivity start to increase dramatically approximately from 10 GHz and no saturation is observed in the measurement range. The real part increases by a factor of around 2 from 1 to 30 GHz. Peit *et al.* found that conductivities at dc and 10 GHz are almost the same.[12] The discrepancy between our data and theirs might be due to the difference of NT sources and purities, which could change the frequency onset of the increase. In many disordered materials the extended pair approximation model $\sigma(\omega) = \sigma_0[1+k(\omega/\omega_0)^s]$, $s<1$, is used to describe the frequency dependence of the conductivity.[4] Using this model to fit the data, we found good agreement with $k=0.134$ and $s \approx 0.55$ for frequency below 10 GHz. At higher frequency, the data begin to deviate from this model, as also reported by Kilbride *et al.* The extended pair approximation model is too simple to fully describe the complete behavior of such a system.[4]

We measured the conductivity of the sample from 173 to 323 K, where shielding materials are normally used. The temperature dependence of Re($\sigma$) is small and gets slightly larger at high frequency, while that of Im($\sigma$) remains small, demonstrating that the shielding effectiveness is not temperature sensitive.

Figure 2(c) shows the dc conductivity and the real part of ac conductivity versus temperature at various fixed frequencies. The real part of conductivity increases with temperature and displays a shallow maximum at $T^* \approx 225$ K. This nonmetallic to metallic crossover phenomena was investigated by Fischer *et al.*[19] The shift of $T^*$ to higher temperature is a result of high intertubular coupling, which was discussed by Bekyarova *et al.*[20] The dc resistivity fits very well to the tunneling model $\rho_T = Ae^{-(T_m/T)} + Be^{(T_b/T_s+T)}$ with $T_m = 775.2$ K, $T_b = 30.5$ K, and $T_s = 45.9$ K. This suggests that the carrier transport in the SWCNT films is governed by tunneling through barriers between conducting NTs. The imaginary part of ac conductivity does not change significantly with temperature, as plotted in Fig. 2(d).

The dielectric constants $\varepsilon = \varepsilon_1 + i\varepsilon_2$ of SWCNT films are calculated using $\varepsilon_1 = 1 + [\text{Im}(\sigma)/\omega\varepsilon_0]$ and $\varepsilon_2 = [\text{Re}(\sigma)/\omega\varepsilon_0]$. As plotted in Fig. 3(a), the real part increases from around $-10^7$ at 10 MHz to around $-10^{-5}$ at 10 GHz. The negative value of $\varepsilon_1$ indicates that the charge transport is dominated by delocalized carriers (metallic).[11] The imaginary part decreases significantly from around $4 \times 10^8$ in the megahertz range to $10^5$ in the gigahertz range.

The shielding effectiveness as a function of frequency for different film thicknesses is calculated through the formula[21]

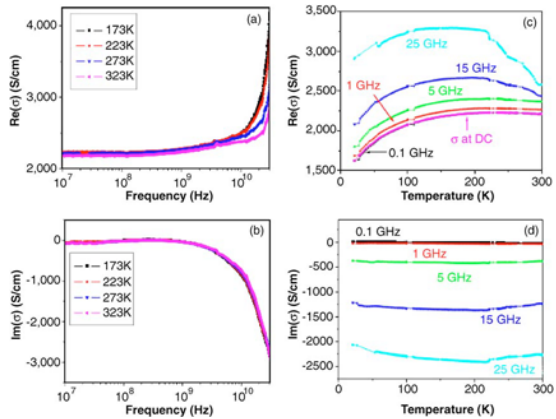

FIG. 2. (Color online) (a) and (b) are real and imaginary parts of conductivity vs microwave frequency at different temperatures; (c) and (d) are real and imaginary parts of conductivity vs temperature at various frequencies, respectively.

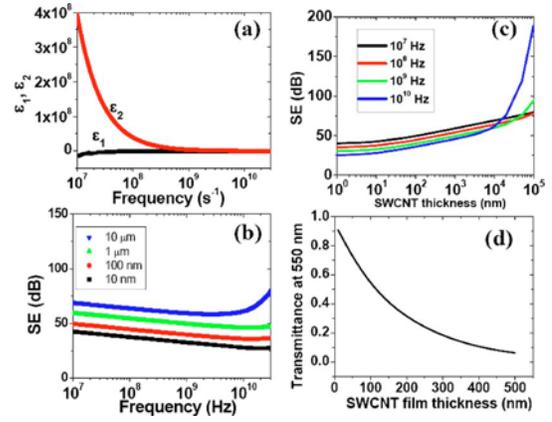

FIG. 3. (Color online) (a) $\epsilon_1$ and $\epsilon_2$ vs frequency, (b) microwave shielding effectiveness (SE) vs frequency, (c) microwave shielding effectiveness vs thickness, and (d) calculated SWCNT optical transmittance vs thickness.

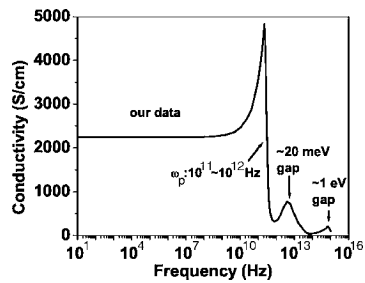

FIG. 4. Sketch of conductivity of SWCNT film in a broad frequency range from dc to visible.

$$SE_{total} = -10 \log T, \quad (2)$$

where $T$ is the electromagnetic radiation transmittance, which depends on the complex index of reflection $n+ik$ and the bulk reflectivity $R$. This formula is applicable when the film thickness is much less than the wavelength. Calculated values of SE deduced from data are plotted in Fig. 3(b). The shielding effectiveness decreases as frequency increases [approximately $SE \propto \log(1/f)$], except for the 10 $\mu$m thickness film for which a sharp increase occurs at about 1 GHz due to internal interference. For the 10 nm film, the shielding effectiveness varies from 43 to 28 dB in the range of 10 MHz–30 GHz. For the 10 $\mu$m film the shielding effectiveness are 70 dB at 10 MHz and 78 dB at 30 GHz . The data for shielding effectiveness versus film thickness at different frequencies are plotted in Fig. 3(c). At all frequencies, the shielding effectiveness increases with thickness [approximately $SE \propto \log(d)$] except for the sharp jump at 10 $\mu$m for 1 and 10 GHz.

SWCNT films are widely studied for transparent electrodes due to the high transmittance in the visible and infrared ranges.[22] For the sake of transparent shielding applications, we calculate the visible transmittance (at a wavelength of 550 nm) versus film thickness using the formula in Ref. 21 with an index of refraction $N=1$ for air and $N=1.06 +0.24i$ for the NT film.[8] In Fig. 3(d), the transmittance of SWCNT films is 90% for 10 nm thickness and 60% for 80 nm thickness. For the 30 nm thickness film, the optical transmittance is about 80% and the shielding effectiveness are 33 dB at 10 GHz, 36 dB at 1 GHz, and 46 dB at 10 MHz. The widely used transparent shielding material, ITO with sheet resistance of 100 $\Omega/\square$, has 30 dB shielding effectiveness at 1 GHz for 80% transmittance with 1 $\mu$m thickness, which is lower than that of SWCNT films.

The conductivity of NT films has been measured by others in different frequency ranges. The frequency dependent conductivity follows the plasma behavior similar to that in metals with a plasma frequency between 0.1 and 1 THz.[11] At higher frequency it was found that the conductivity has two peaks. One peak is near 10–20 meV corresponding to the secondary band gap which may be caused by rolling up graphene to create the NT.[6] The location of this peak varies substantially from sample to sample. The other peak is the optical band gap around 1 eV. The conductivity at the optical band gap edge slightly depends on chemical doping and NT purity, whereas that of the secondary band gap largely depends on doping and purity.[8] Combined with our measurement, a master sketch of conductivity is made in Fig. 4. For different NT sources, purities, and chemical doping levels, the position and height of the peak corresponding to the 20 meV gap as well as the plasma frequency will change. The peak at the optical band gap remains the same.

We found that the shielding effectiveness of SWCNT films satisfies requirements for some commercial applications (e.g., cell phones require approximately 20 dB shielding effect). For some extremely high shielding requirements, such as for magnetic resonant imaging window where 60 dB shielding effectiveness is required, the current transparent NT films still need to be improved. This can be done by using optimized chemical doping, longer tubes, or solely-metallic NTs, which will increase the shielding effectiveness without sacrificing the optical transmittance.[23] Along with more development of commercial NT sources, SWCNT films are promising for transparent EMI shielding applications. The master sketch of conductivity provides guidance for exploring the potential applications of NT films as a type of electrical and photonic material.

The authors thank Degiorgi *et al.* for providing the original optical data of Ref. 8 This work has been supported by NSF Grant Nos. (DMR-0404029) (DMR-0322844), and (DMR-0302596).